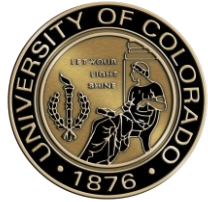

# Choosing the Right Communication Protocol for your Web Application

Literature review by: **Mohamed Hassan**[*1]
*grad student*
*Dept. of computer science*
Mohamed.hassan@colorado.edu
https://orcid.org/
https://scholar.google.com/

[1]*College of Engineering and Applied Sciences, University of Colorado, Boulder, CO, United States of America

## I. KEYWORDS

HTTP/1.1 (hypertext transfer protocol) [1], [2], [3],
HTTP/2 (hypertext transfer protocol) [1], [2], [3], [4], [5],
HTTP/3 (hypertext transfer protocol) [1], [2], [3], [3], [6],
Application Programming Interfaces (APIs) [7], [8],
Soap (Simple Object Access Protocol) [9],
GraphQL (Graph Query Language) [10],
gRPC (Remote Procedure Call) [11], [12],
WebSockets [13],
SSE (Server-Sent Events) [8],
MQTT (Message Queuing Telemetry Transport) [14],
IDL (Interface Definition Language) [15],
CRUD (Create, Read, Update, Delete) [16],
QUIC (Quick UDP[17] Internet Connections, pronounced quick) [18],
Server [19],
Protocol [20],
Client [21], [22],
Real time [23],
Latency [24],
Request [25],
Networks [26],
Bandwidth [27],
Unidirectional [28],
Query [29],
Protobuf (Protocol Buffers) [30],
Interface [31],
Process [32],
Multiplexing [33],
Endpoint [34],
TCP [35], [36],
UDP [17],
IoT (Internet of Things) [37],
Clients and servers [38],
JSON [39],
XML [40], [41],
SMTP [42],
Microservice [43] Author published paper [44],
Full-duplex communication [45],
TCMTF (multiplexing, header compression) [46],

## II. ABSTRACT

Selecting the appropriate communication protocol is crucial for optimizing the performance, scalability, and user experience of web applications. In the diverse ecosystem of web technologies, various protocols [20] like RESTful APIs [7], [8], gRPC [11], [12], WebSockets [13], and others serve distinct purposes. RESTful APIs [7], [8] are widely favored for their simplicity and stateless nature, making them ideal for standard CRUD [16] operations. They offer a straightforward approach to interacting with resources over HTTP/1.1 [1], [3], [6], providing broad compatibility and ease of integration across different platforms. However, in scenarios where applications require high efficiency and real-time [23] communication, gRPC [11], [12] and WebSockets [13] emerge as powerful alternatives.

Each protocol comes with its strengths and limitations, influencing factors such as ease of implementation, performance under load, and support for complex data structures. RESTful APIs [7], [8], while easy to use and widely supported, may introduce overhead due to their stateless nature and reliance on multiple HTTP/1.1 [1], [2], [3] requests [25]. In contrast, gRPC [11], [12] advanced features, while powerful, require a steeper learning curve and more sophisticated infrastructure. Similarly, WebSockets [13], while excellent for real-time applications [23], require careful management of persistent connections and security considerations.

This paper explores the key considerations in choosing the right communication protocol, emphasizing the need to align technical choices with application requirements and user expectations. By understanding the unique attributes of each protocol, developers can make informed decisions that enhance the responsiveness and reliability of their web applications. The choice of protocol can significantly impact the user experience, scalability, and maintainability of the application, making it a critical decision in the web development process.

## III. INTRODUCTION

In the evolving landscape of web development, efficient and effective communication between clients and servers [38] is essential for creating robust, scalable, and responsive web applications. As web applications grow in complexity, the demand for faster, more reliable, and more dynamic communication methods has increased. This need has driven the development of various communication protocols [20] and technologies, each tailored to address specific challenges and use cases within web applications. From handling basic data retrieval tasks to enabling sophisticated real-time [23] interactions, these communication methods are the backbone of modern web application architecture.

Over the years, developers have moved from simple request-response models to more complex and capable communication protocols [20]. Traditional methods like RESTful APIs [7], [8] have provided a solid foundation for interacting with resources over the web, utilizing the familiar HTTP/1.1 [1], [2], [3] protocol to send and receive data in a stateless manner. RESTful's simplicity, broad compatibility, and ease of integration have made it the go-to solution for many applications, particularly those focused on standard CRUD (Create, Read, Update, Delete) [16] operations. However, as the need for more dynamic and interactive applications has grown, the limitations of RESTful APIs [7], [8], such as their overhead and inefficiency in handling real-time [23] updates, have become more apparent.

To address these limitations, more advanced communication technologies like gRPC [11], [12] and WebSockets [13] have been developed. gRPC [11], [12], an open-source framework originally developed by Google, leverages the power of HTTP/2 [1], [2], [3], [4], [5] and Protocol Buffers [20], [30] to enable high-performance, low-latency communication. Its ability to handle bi-directional streaming and its strong typing through IDL (Interface Definition Language) [15] make it particularly well-suited for microservices architectures [43], where efficient and reliable inter-service communication is critical.

WebSockets [13], on the other hand, provide a solution for real-time [23], full-duplex communication [45] between clients and servers [21], [22], [38]. Unlike the traditional request-response model, WebSockets [13] maintain an open connection that allows data to flow freely in both directions, enabling applications like live chat, online gaming, and collaborative platforms to operate with minimal latency. This capability is essential in environments where immediate feedback and interaction are required, as it allows for a seamless and responsive user experience.

In addition to these, other communication technologies like SSE (Server-Sent Events) [8], MQTT (Message Queuing Telemetry Transport) [14], and HTTP/2/3 [1], [2], [3], [4], [5] have also emerged, each catering to specific needs within the web development ecosystem. SSE [8], for instance, is ideal for applications that require continuous updates from the server, such as news feeds or live sports scores, while MQTT [14] is commonly used in IoT (Internet of Things) [37] environments where low-bandwidth and high-latency communication are common challenges.

Whether the goal is to optimize for speed, scalability, security, or user experience, understanding the strengths and weaknesses of each communication method is crucial for building applications that meet the demands of today's fast-paced, interconnected world.

The requirement to rebuild and relaunch the complete application for every modification further restricts the development process's flexibility and effectiveness.

It also presents difficulties for vertically (adding more resources) [47] or horizontally (adding more instances) [47] expanding the program, since it could necessitate large architectural or infrastructure changes.

The microservice's primary [43] advantages are its scalability, agility, and modularity. It enables the application to be divided into more manageable, smaller components that may be independently created, deployed, and scaled.

It also makes it possible for new features and upgrades to be delivered more quickly and often, as well as for new tools and technologies to be integrated more easily. The microservice [43] concept does, however, have many shortcomings, including complexity, inconsistent behavior, and lack of stability.

More advanced testing and monitoring instruments, as well as increased unit cooperation and communication, are needed. In addition, it adds more overhead for data transfer and network traffic, as well as additional potential points of failure and performance problems.

# IV. API Application Programming Interfaces

An API (Application Programming Interface) [7] is a set of rules and protocols [20] that allow one piece of software to interact with another. In the context of web applications, APIs [7], [8] enable communication between different software components, allowing them to exchange data and perform various operations. APIs [7], [8] are typically categorized into several types, including RESTful APIs, SOAP [9] APIs [7], [8], and GraphQL [10], each with its unique features and use cases.

## 1.1. RESTful APIs

RESTful (Representational State Transfer) [7], [8] is an architectural style that defines a set of constraints for creating web services. RESTful APIs [7], [8] are the most common type of API used in web development. They rely on stateless communication, meaning each request [25] from the client [21], [22] to the server [19] must contain all the information needed to understand and process the request [25].

### Characteristics:

- Statelessness: Each request [25] is independent and contains all necessary information.
- Scalability: Due to their stateless nature, RESTful APIs [7], [8] can easily scale.
- Uniform Interface: RESTful APIs [7], [8] use standard HTTP/1.1 [1], [2], [3] methods (GET, POST, PUT, DELETE) to perform operations.

### Use Cases:

- RESTful APIs [7], [8] are widely used for web services that require CRUD (Create, Read, Update, Delete) [16] operations.
- They are suitable for applications where resource representations (such as JSON [39] or XML [40]) are exchanged over HTTP/1.1 [1], [2], [3].

### Advantages:

- Simplicity and ease of use.
- Widespread adoption and support across various platforms and languages.

### Limitations:

- Lack of flexibility in defining complex queries.
- Overhead of HTTP/1.1 [1], [2], [3] requests [25] can be inefficient for real-time [23] applications.

## 1.2. SOAP

SOAP (Simple Object Access Protocol) [9] is a protocol for exchanging structured information in web services. Unlike RESTful, which is more flexible and relies on standard HTTP/1.1 [1], [2], [3] methods, SOAP [9] is a protocol with strict rules and is often used in enterprise-level applications that require high security and reliability.

### Characteristics

- Protocol-based: SOAP [9] uses XML [40], [41] as its message format and typically relies on HTTP/1.1 [1], [2], [3], SMTP [42], or other protocols [20] for message negotiation and transmission.
- Extensibility: SOAP [9] supports WS-* [48] standards, which offer additional features like security (WS-Security) [48], transactions, and reliable messaging.

### Use Cases:

SOAP [9] is used in scenarios where security and transactional reliability are critical, such as in financial services, telecommunications, and healthcare.

### Advantages:

- Built-in error handling and retry logic.
- Strong standards for security and compliance.

### Limitations:

- Verbosity of XML can lead to larger message sizes.
- Complexity in implementation and maintenance compared to RESTful.

## 1.3. GraphQL

GraphQL [10] is a query language for APIs [7], [8] and a runtime for executing those queries. Developed by Facebook, it allows clients [21], [22] to request [25] exactly the data they need, making it a powerful alternative to RESTful.

### Characteristics:

- Query-based: Clients [21], [22] define the structure of the response, ensuring that only the required data is retrieved.
- Single Endpoint [34]: Unlike RESTful, which uses multiple endpoints [34], GraphQL [10] typically uses a single endpoint [34] for all operations.

### Use Cases:

- Applications that require a highly customizable and efficient data retrieval process.

- Scenarios where minimizing over-fetching or under-fetching of data is crucial.

**Advantages:**

- Efficient data retrieval with minimal bandwidth usage.
- Flexibility in querying complex data structures.

**Limitations:**

- Steeper learning curve for developers unfamiliar with the query language.
- Potential performance issues with deeply nested queries.

### 1.4. gRPC

**Overview**

gRPC [11], [12] is an open-source RPC (Remote Procedure Call) framework developed by Google. It uses HTTP/2 [1], [2], [3], [4], [5] for transport, Protobuf (Protocol Buffers) [20], [30] as its IDL (Interface Definition Language) [15], and offers features like authentication, load balancing, and more.

**Characteristics:**

- Protobuf: gRPC [11], [12] uses Protobuf (Protocol Buffers) [20], [30] for serializing structured data, which is more efficient than JSON [39] or XML [41].
- HTTP/2 [1], [2], [3], [4], [5]: gRPC [11], [12] benefits from HTTP/2's [1], [2], [3], [4], [5] features like multiplexing, header compression [46],, and low latency.

**Use Cases:**

- gRPC [11], [12] is ideal for microservices architecture [43], where different services need to communicate with each other efficiently.
- It's also used in real-time [23] communication systems, such as video conferencing or online gaming.

**Advantages:**

- High performance due to the use of Protobuf (Protocol Buffers) [20], [30] and HTTP/2 [1], [2], [3], [4], [5].
- Strongly-typed contracts between client [21], [22] and server [19], reducing runtime errors.

**Limitations:**

- Limited support for some web browsers, making it less suitable for client-side web applications.

- More complex setup and tooling compared to RESTful APIs [7], [8].

### 1.5. WebSockets

**Overview**

WebSocket [13] is a protocol providing full-duplex communication [45] channels over a single, long-lived connection. It allows for real-time [23] data exchange between a client and a server with minimal overhead.

**Characteristics**:

- Full-duplex communication [45]: WebSocket enables bidirectional communication, allowing the client and server to send messages to each other independently.
- Persistent Connection: Unlike HTTP/1.1 [1], [2], [3], which requires a new connection for each request/response, WebSocket maintains a single connection open for the entire session.

**Use Cases**:

- Real-time [23] applications like chat apps, live updates, and multiplayer online games.
- Scenarios where low latency is critical, such as financial trading platforms or IoT [37] devices.

**Advantages**:

- Efficient use of bandwidth due to reduced HTTP/1.1 [1], [2], [3] headers.
- Real-time [23] communication with minimal latency.

**Limitations**:

- Not suitable for applications where only occasional communication is needed.
- Requires additional security considerations due to the persistent connection.

### 1.6. Server-Sent Events (SSE) [8]

**Overview**

SSE (Server-Sent Events) [8] is a standard allowing servers to push updates to clients over a single HTTP/1.1 [1], [2], [3] connection. Unlike WebSockets [13], SSE [8] is unidirectional, meaning only the server can send data to the client.

**Characteristics**:

- Unidirectional Communication: The server can send continuous updates to the client, but the client cannot send data back over the same connection.
- Text-based Protocol: SSE[8] uses simple HTTP/1.1 [1], [2], [3] and sends data in text format.

Use Cases:

- Applications requiring continuous updates from the server, such as news feeds, social media notifications, or live sports scores.
- Scenarios where the server needs to push updates, but client interaction is minimal.

Advantages:

- Simpler to implement compared to WebSockets [13].
- Built-in support in most modern web browsers.

Limitations:

- Limited to unidirectional communication.
- Not suitable for complex real-time [23] interactions that require client-to-server communication.

### 1.7. HTTP/2 and HTTP/3

Overview

HTTP/2 [1], [2], [3], [4], [5] and HTTP/3 [1], [2], [3], [3], [6] are the latest versions of the HTTP/1.1 [1], [2], [3] protocol, designed to improve performance and efficiency in web communication. They introduce several features that enhance data transfer speeds and reduce latency.

Characteristics of HTTP/2 [1], [2], [3], [4], [5]:

- Multiplexing: Multiple requests [25] and responses can be sent simultaneously over a single connection.
- Header Compression: Reduces the size of HTTP/1.1 [1], [2], [3] headers, improving performance.
- Server Push: Allows the server to send resources to the client before they are requested.

**Characteristics** of HTTP/3 [1], [2], [3], [3], [6]:

- QUIC Protocol [18]: HTTP/3 [1], [2], [3], [3], [6] is built on the QUIC protocol [18], which uses UDP [17] instead of TCP [35], [36], offering faster connection establishment and improved performance over lossy networks.
- Improved Latency: QUIC's design reduces latency, especially in mobile and wireless networks.

Use Cases:

- HTTP/2 [1], [2], [3], [4], [5] and HTTP/3 [1], [2], [3], [3], [6] are used in modern web applications to improve page load times and overall performance.
- They are particularly beneficial in scenarios with multiple small requests, such as loading a webpage with many assets.

Advantages:

- Significant performance improvements over HTTP/1.1 [1], [2], [3].
- Better handling of network congestion and packet loss.

Limitations:

- Requires support from both client and server, and some legacy systems may not fully support these protocols[20].
- Complexity in implementation, particularly for HTTP/3 [1], [2], [3], [3], [6] and QUIC [18].

### 1.8. MQTT (Message Queuing Telemetry Transport)

Overview

MQTT [14] is a lightweight messaging protocol designed for low-bandwidth, high-latency, or unreliable networks. It is widely used in IoT (Internet of Things) [37] applications, where devices need to communicate with each other or a central server.

Characteristics:

- Publish/Subscribe Model: MQTT [14] uses a broker to manage communication between clients, allowing for efficient message distribution.
- Low Overhead: The protocol is designed to minimize network bandwidth and device resource usage.

Use Cases:

- IoT [37] applications, where devices with limited resources need to communicate over constrained networks.
- Scenarios requiring real-time [23] messaging with minimal

### V. CONCLUSION

Selecting the appropriate communication protocol for a web application is a critical decision that can have far-reaching effects on the application's overall performance, scalability, and user experience. Given the diverse landscape of communication protocols—ranging from the simplicity of RESTful APIs to the advanced features of gRPC [11], [12], WebSockets [13], and others—developers must carefully evaluate the specific needs of their application to determine the best fit. RESTful APIs have long been the go-to solution for applications requiring basic, stateless operations, due to their wide adoption, simplicity, and ease of integration. However, when an application requires higher performance, such as low-latency responses or real-time bidirectional communication, protocols like gRPC [11], [12] and WebSockets [13] offer more robust solutions. gRPC [11], [12], with its ability to efficiently serialize data using Protobuf (Protocol Buffers) [30], provides a streamlined and high-performance alternative, especially suited for microservices architectures.

WebSockets [13], on the other hand, are ideal for real-time applications, such as live streaming or collaborative tools, where continuous, bidirectional communication between the client and server is essential for maintaining a smooth user experience. Additionally, other specialized protocols like SSE (Server-Sent Events) [8] and MQTT [14] have emerged to address specific use cases. SSE [8], for instance, is designed for scenarios that require continuous updates from the server, such as real-time notifications or data feeds, while MQTT [14] is commonly used in IoT [37] environments due to its ability to handle low-bandwidth, high-latency connections efficiently.

Ultimately, the decision of which communication protocol to use must align with the application's technical requirements, user interaction patterns, and long-term goals. Developers need to consider factors like data structure complexity, the need for real-time interactivity, scalability, and overall efficiency.

Choosing the right protocol can significantly enhance the responsiveness and reliability of the application, leading to improved user satisfaction and better scalability as the application grows.

Furthermore, as web technologies continue to evolve, it is essential for developers to stay informed about advancements in communication protocols to make more strategic decisions that future-proof their applications. By keeping up-to-date with the latest developments and understanding the trade-offs between different communication methods, developers can ensure they build applications that are not only efficient and high-performing but also adaptable to changing requirements and emerging technologies.

In a world where web applications are increasingly expected to provide real-time, seamless interactions and scale to meet growing user demands, making an informed choice about communication protocols is essential for the long-term success and sustainability of modern web applications.

## VI. ACKNOWLEDGMENTS

## VII. LLM DECLARATION

The author, being a non-native English speaker, utilized the GPT-4 model as a refinement instrument in the crafting of this review paper.

- Rectify grammatical errors and misspellings.
- Maintain linguistic uniformity and fluency.